\pdfoutput=1
\documentclass[preprint,12pt,preprintnumbers,amsmath,amssymb,floatfix,endfloats*]{revtex4}
\usepackage{graphicx}% Include figure files
\usepackage{dcolumn}% Align table columns on decimal point
\usepackage{bm}% bold math
\usepackage{amsfonts}

\DeclareMathAlphabet{\mathsfsl}{OT1}{cmr}{bx}{it}
\begin{document}
%----------------------------------------------------------------------%
% Title
%----------------------------------------------------------------------%
%
\title{The critical pressure for microfiltration of oil-in-water emulsions using slotted-pore membranes}
\author{Tohid Darvishzadeh$^{1}$, Bishal Bhattarai$^{2}$, Nikolai V. Priezjev$^{2,3}$}
\affiliation{$^{1}$Department of Mechanical Engineering, Michigan
State University, East Lansing, MI 48824}
\affiliation{$^{2}$Department of Mechanical and Materials
Engineering, Wright State University, Dayton, OH 45435}
\affiliation{$^{3}$National Research University Higher School of
Economics, Moscow 101000, Russia}
\date{\today}
%
%----------------------------------------------------------------------%
%\twocolumn[
%----------------------------------------------------------------------%
% Abstract
%----------------------------------------------------------------------%
\begin{abstract}

The influence of geometrical parameters and fluid properties on the
critical pressure of permeation of an oil micro-droplet into a
slotted pore is studied numerically by solving the Navier-Stokes
equations. We consider a long slotted pore, which is partially
blocked by the oil droplet but allows a finite permeate flux. An
analytical estimate of the critical permeation pressure is obtained
from a force balance model that involves the drag force from the
flow around the droplet and surface tension forces as well as the
pressure variation inside the pore. It was found that numerical
results for the critical pressure as a function of the oil-to-water
viscosity ratio, surface tension coefficient, contact angle, and
droplet radius agree well with theoretical predictions.  Our results
show that the critical permeation pressure depends linearly on the
surface tension coefficient, while the critical pressure nearly
saturates at sufficiently large values of the viscosity ratio or the
droplet radius.  These findings are important for an optimal design
and enhanced performance of microfiltration systems with slotted
pores.

\vskip 0.1in


\vskip 0.1in

\end{abstract}

\maketitle

\section{Introduction}

The efficient microfiltration of oil-in-water emulsions, that are
commonly found in produced water from oil/gas recovery or as a
byproduct of metal finishing processes, is important for protection
of the aquatic environment and recovery of clean
water~\cite{Tian18,Jassby17}.  The advantages of the membrane
filtration methods, as compared to sedimentation, dissolved gas
flotation and centrifugation, include relatively low maintenance
cost, mechanical stability and simple operating
conditions~\cite{Zhu14,Dong14,WangPan15,Meynen16}.   The membrane
separation process is based on rejection of nonwetting oil droplets
that are larger than membrane pores, while at the same time allowing
permeate flux, which is induced by the applied pressure across the
membrane. However, the major practical challenge in achieving a high
permeate flux is the membrane fouling~\cite{Song98}.  Common
antifouling strategies include crossflow filtration, physical or
chemical cleaning, surface wettability modification, stabilization
by surfactants, and electrostatic
repulsion~\cite{Ma15,Freeman17,Ramon17,Beatriz17,Rahaman17,Yu18}.
More recently, a multicontinuum approach for estimation of the
permeation capacity of thin flat membranes with a broad distribution
of pore openings was formulated for oil-water emulsions with
different droplet sizes and then validated using the experimental
data~\cite{Salama17,Salama18}.

\vskip 0.05in

At the microscopic level, the critical transmembrane pressure
required for an oil droplet to permeate into a membrane pore can be
deduced from the analysis of the Young-Laplace pressure across
curved oil-water interfaces inside and above the
pore~\cite{Nazzal96}.   In the absence of crossflow, a simple model
for the critical pressure of permeation of an oil droplet into a
circular pore was validated via detailed numerical
simulations~\cite{Tohid12} and experimental
measurements~\cite{Nazzal96}.  As an aside, an analytical expression
for the critical permeation pressure in the case of a continuous oil
film above a membrane surface with a pore of arbitrary cross-section
was obtained and validated for rectangular and elliptical
pores~\cite{Tohid12}.   It was later shown that in the presence of
crossflow along the membrane surface, the critical pressure
increases due to the drag force generated by the shear flow around
an oil droplet, and, at sufficiently high shear rates, the droplet
above the pore breaks up into two
segments~\cite{Tohid12,Tohid13,Maryam16}. The results of numerical
simulations have demonstrated that the breakup capillary number and
the increase in critical pressure due to crossflow are nearly
independent of the contact angle but depend strongly on the
oil-to-water viscosity ratio, surface tension, and drop-to-pore size
ratio~\cite{Tohid13}. Recent experimental studies on crossflow
microfiltration have shown that membrane fouling might involve
several stages, namely, droplet attachment and clustering, droplet
deformation and coalescence~\cite{Emily16} and that the critical
flux decreases with salt concentration~\cite{Tarabara17}.  Using CFD
modeling, it was also recently found that membrane fouling can be
reduced by applying sufficiently large electric field, which results
in the oil droplet detachment from the membrane
surface~\cite{Maryam17}.

\vskip 0.05in

One of the key factors for the efficient microfiltration of
oil-in-water dispersions is the pore size and
shape~\cite{Apel01,Ziel08,Ullah11,Fan16,Vernerey16}.   In
particular, it was reported that, under similar operating
conditions, the critical flux through a membrane with slotted pores
is much higher than the flux through circular pore
membranes~\cite{Bromley02}.  This result can be intuitively
understood from the fact that an oil droplet can only partially
block a long slotted pore and thus allow a finite permeate flux,
whereas a relatively large oil droplet can completely block a
circular pore.   It was later shown that the experimental data for
rejection of oil droplets by slotted pore membranes are well
described by theoretical predictions, assuming that a spherical
droplet deforms into an ellipsoid inside the pore and taking into
account the static and drag forces~\cite{Ullah12a}.  An improved
rejection of oil droplets through the slotted pore membrane can also
be achieved by imposing vibration that leads to shear-induced
migration and an inertial lift of droplets from the membrane
surface~\cite{Ullah12b}. Despite significant progress, however, the
exact physical mechanism of the droplet permeation into a slotted
pore even at the initial filtration stage and in the absence of
crossflow is not fully understood.

\vskip 0.05in

In this paper, we investigate the process of oil droplet permeation
into a slotted pore under applied pressure using numerical
simulations and the Volume of Fluid method to track the oil-water
interface. The critical permeation pressure is first estimated
theoretically using a force balance analysis of the capillary forces
and the drag force from the flow around the droplet.  It will be
shown that theoretical predictions agree well with the numerically
obtained critical pressure as a function of the oil-to-water
viscosity ratio, the surface tension coefficient, the contact angle,
and the droplet radius.

\vskip 0.05in

The reminder of the paper is organized as follows.  In the next
section, the numerical simulation method and governing equations are
presented.  The theoretical prediction for the critical permeation
pressure based on the force balance arguments is given in
Sec.\,\ref{subsec:theory}, and the results of numerical simulations
and comparison with analytical predictions are reported in
Sec.\,\ref{subsec:effect_of_parameters}. The brief summary of the
results is provided in the last section.

\section{Numerical simulations}
\label{sec:numerical_method}

% introduction

The interaction of an oil droplet with the porous membrane surface
was studied numerically by solving the Navier-Stokes equation, which
is implemented in the commercial software ANSYS
FLUENT~\cite{fluent}. In turn, the dynamics of an oil-water
interfaces was tracked by the the Volume of Fluid (VOF) method,
where computational cells contain information on the volume fraction
of each phase~\cite{Hirt81}.    More specifically, we considered an
oil droplet at the entrance of a slotted pore and applied a pressure
difference across the membrane, which induces a flow around the
droplet, as shown schematically in
Fig.\,\ref{fig:schem_force_balance}.   The numerical methodology for
the problem of an oil droplet at a slotted pore is very similar to
the numerical setup used in our previous papers, where the
permeation, rejection and breakup of an oil droplet at a circular
pore was investigated for a number of material parameters and
various operating conditions~\cite{Tohid12,Tohid13}.  In particular,
we performed test simulations to determine the appropriate domain
size and grid resolution necessary to accurately capture the effects
of interface curvature and flow around the droplet. It was found
that the computational domain has to be at least 4 times the size of
the droplet and that at least 20 mesh cells are required across the
pore. As discussed below, we considered an infinitely long slotted
pore by using a symmetry boundary condition, thus eliminating the
finite size effects due to pore ends.   The numerical values of the
geometrical and material parameters are listed at the end of
Sec.\,\ref{subsec:theory}.

\vskip 0.05in

% governing equations

In the Volume of Fluid method~\cite{Hirt81}, the oil-water interface
is specified by the volume fraction $\alpha$, which is coupled to
the flow via the solution of the transport equation as follows:
\begin{equation}
\frac{\partial{\alpha}}{{\partial}{t}}+{\nabla}{\cdot}\,({\alpha}{\textbf{V}})=\,0,
\label{eq:continuity}
\end{equation}
where $\textbf{V}$ is the local velocity vector. The material
properties near interfaces and in the bulk are averaged within each
cell based on the volume fraction of each phase. For example, the
averaged local density is obtained as follows:
\begin{equation}
\rho = \alpha\,\rho_{2}+(1-\alpha)\,\rho_{1}.
\label{eq:compatibility}
\end{equation}
Next, the following momentum equation is solved numerically using
the average fluid viscosity and density:
\begin{equation}
\frac{\partial}{{\partial}{t}}({\rho}{\textbf{V}})+\nabla\cdot(\rho{\textbf{V}}{\textbf{V}})
=-\nabla{p}+\nabla\cdot[\mu(\nabla{\textbf{V}}+\nabla{\textbf{V}}^{T})]+\rho\,\textbf{g}+\textbf{F},
\label{eq:momentum}
\end{equation}
where $\textbf{V}$ is the velocity vector, $\textbf{g}$ is gravity,
and $\textbf{F}$ is the local surface tension force at the curved
interface, which is defined as:
\begin{equation}
\textbf{F} =
\sigma\,\frac{\rho\,\kappa\nabla\alpha}{\frac{1}{2}(\rho_{1}+\rho_{2})},
\label{eq:surface_tension}
\end{equation}
where $\kappa$ is the local curvature of the oil-water interface and
$\sigma$ is the surface tension coefficient. The local curvature,
$\kappa$, is computed as follows:
\begin{equation}
\kappa=\,\frac{1}{|\textbf{n}|}\Big[\Big(\frac{\textbf{n}}{|\textbf{n}|}\cdot\,\nabla\Big)|\textbf{n}|-
(\nabla\cdot\,\textbf{n})\Big],
\label{eq:curvature}
\end{equation}
where $\textbf{n}$ is the unit vector normal to the interface. Thus,
the surface tension force given by Eq.\,(\ref{eq:surface_tension})
acts in the direction normal to the interface, and the magnitude of
the force is larger for more curved interfaces~\cite{Gerlach06}.
Finally, the local orientation of the oil-water interface at the
membrane or pore surfaces is determined by the static contact
angle~\cite{Brackbill92}. In practice, the unit vector normal to the
interface at the solid surface is estimated as follows:
\begin{equation}
\textbf{n}_{i}=\,\textbf{n}_{w}\,\textrm{cos}\,\theta+\,\textbf{n}_{t}\,\textrm{sin}\,\theta,
\label{eq:contact_angle}
\end{equation}
where the unit vectors $\textbf{n}_{w}$ and $\textbf{n}_{t}$ are
normal to the wall and normal to the contact line, respectively. In
our simulations, the contact angle is constant and it is strictly
imposed at every surface of the membrane including the interior of
the pore. Therefore, the contact angle hysteresis and dynamic
contact angle effects were not considered in the present study.
Also, the the gravitational and electrostatic forces were not
included in the analysis.

\vskip 0.05in

% Q: How is a simulation run (how are the points on the graphs acquired)?

The solver starts by generating a spherical droplet right above the
pore entrance via a User-Defined-Function (UDF). Based on the
droplet size and pore width, the UDF positions the droplet so that
it barely touches the edges of the pore, thus facilitating
attachment to the membrane surface and preventing upward migration
at low transmembrane pressures.  Another UDF subroutine applies a
pressure-outlet boundary condition at the bottom of the pore, and
the pressure difference across the simulation cell is increased from
an initial value to a final value in a certain number of steps.
During each step, the solver is iterated many times to reach a
steady state. The position of the droplet interface inside the pore
is actively monitored. With each increment of the transmembrane
pressure up to a critical value, the droplet interface inside the
pore is displaced away from the edges, while the whole droplet
remains pinned at the pore entrance. At a particular step, a slight
increase in the transmembrane pressure results in a dramatic
acceleration of the droplet interface inside the pore. This
indicates the droplet permeation, and the critical pressure is
calculated as the average value of the pressure jump at that
particular step with an error bar of the step size.  To compute the
critical permeation pressure more accurately, the same process was
repeated multiple times.

% Every increase in the transmembrane pressure results in a small
% increase in the vertical position of the interface with respect to the pore edge.

\section{Results}
\label{sec:Results}

\subsection{An estimate of the critical pressure based on the force balance}
\label{subsec:theory}

% The droplet is assumed to have slowly deposited on the pore and that
% impact forces are not significant. The contact angle is constant and
% strictly imposed on every surface of the membrane including the
% \textbf{interior} inside of the pore. Therefore, contact angle
% hysteresis and dynamic contact angle effects are not considered.
% Since the interface is static for the most part until permeation,
% the motion of the contact line is not a primary concern. Since the
% Reynolds number is very low, the flow is assumed to be laminar.
% Since the variations in temperature are neglected, the viscosity of
% each phase is constant throughout every simulation.

% Q: Why do you pick this specific control volume for force balance study?

\begin{figure}[t]
\includegraphics[width=13.0cm,angle=0]{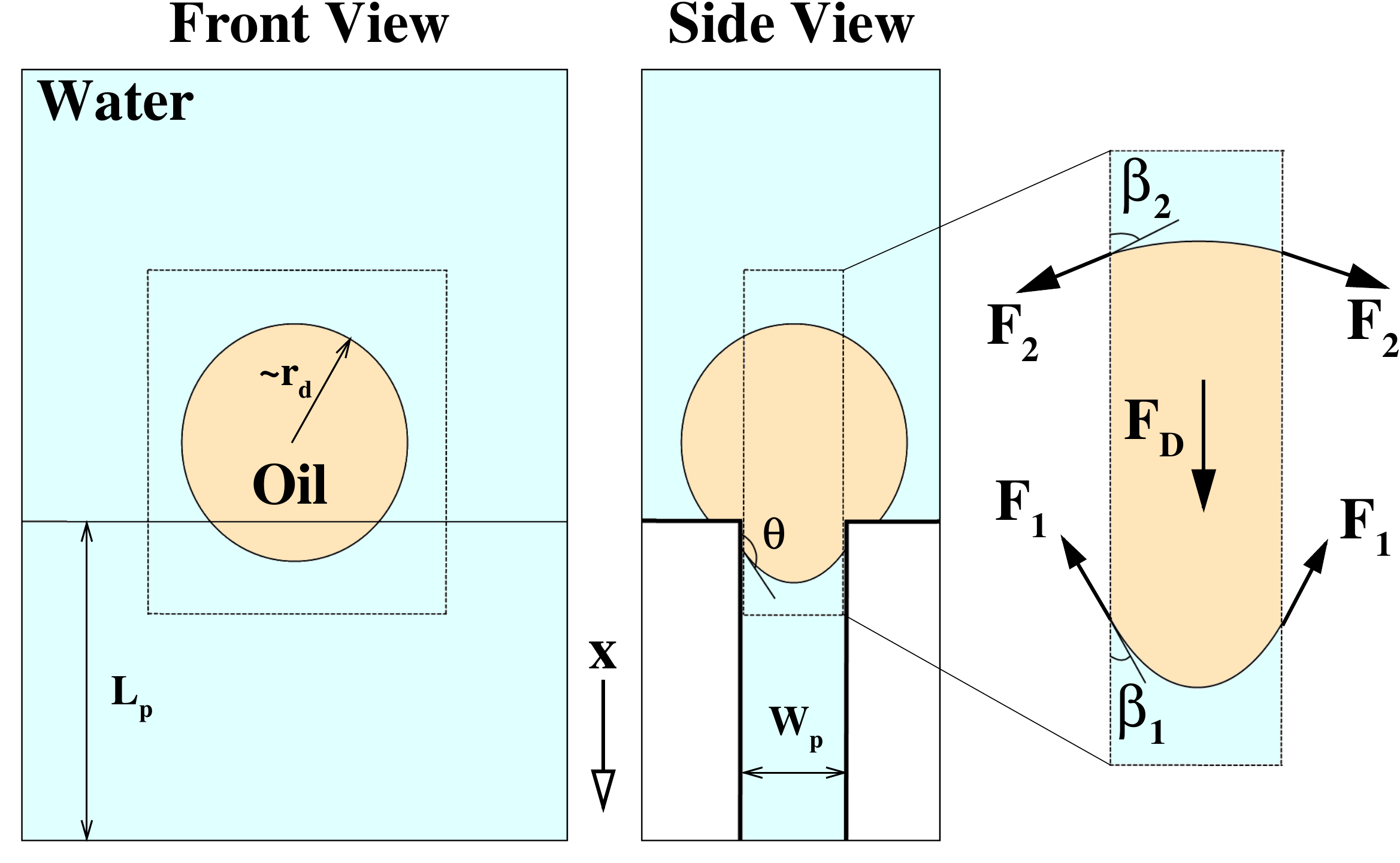}
\caption{(Color online) A schematic illustration of the oil droplet
residing at the entrance of a slotted pore.  The parameters $L_p$
and $W_p$ denote the length and width of the pore. The control
volume is indicated by the dashed rectangles.  An enlarged view of
the control volume on the right-hand side shows the surface tension
and drag forces as well as the local contact angles. }
\label{fig:schem_force_balance}
\end{figure}

We consider an oil droplet residing at the entrance of a slotted
pore shown schematically in Fig.\,\ref{fig:schem_force_balance}.
There are two main forces acting on the droplet; namely, the drag
force from the viscous flow and the capillary forces. The
competition between these forces determines the critical pressure
required for permeation of the droplet inside the slotted pore. The
permeation of the droplet involves a displacement of its interface
into the pore, which, due to symmetry, moves along the centerline in
the $x$ direction (see Fig.\,\ref{fig:schem_force_balance}).
Therefore, only the $x$ component of the forces needs to be included
in the analysis and the other components are canceled due to
symmetry. We define a control volume around the droplet as shown in
Fig.\,\ref{fig:schem_force_balance}. The viscous and pressure forces
acting outside the control volume on two sides of the droplet (the
side view) are counter-balanced by forces from pressure and surface
tension forces from the surface of the membrane.  These forces are
expected to be much smaller than the forces acting on the control
volume, since the shear stresses are smaller in the region outside
the control volume.

% CHECK NUMERICALLY ALL THE FORCES ACTING ON EACH FACE OF YOUR CONTROL
% VOLUME AND USE THAT TO PROVE THAT THE FORCES ACTING ON SOME FACES ARE NEGLIGIBLE

\vskip 0.05in

First, the viscous flow around the droplet exerts a drag force,
which is proportional to the viscosity of the continuous phase,
viscosity ratio, average flow velocity, and droplet size. Since the
flow field around the droplet gradually varies near the pore
entrance, an exact analytical expression for the drag force is not
available. However, due to the linearity of the flow at small
Reynolds numbers, it can be assumed that the drag force is linearly
proportional to the following parameters:
\begin{equation}
F_{D}\propto\,f(\lambda)\,\mu_{w}\,\bar{u}\,r_{d},
\label{eq:drag_force}
\end{equation}
where $\lambda$ is the ratio of oil and water viscosities, $\mu_{w}$
is the viscosity of water, and $\bar{u}$ is the average flow
velocity around the droplet. The drag force is also a function of
the viscosity ratio via $f(\lambda)$. The form of such function is
not readily known for our geometry. The flow accelerates at the
entrance of the pore, while inside the pore and far away from the
droplet, the velocity remains constant. This velocity is a result of
the pressure gradient inside the pore. The pressure gradient induces
a planar Hagen-Poiseuille flow inside the pore with a parabolic
velocity profile.  The average flow velocity in the pore is thus
proportional to the following combination:
\begin{equation}
\bar{u}\propto\,\frac{{W_{p}}^{2}\,|\Delta{P}|}{\mu_{w}\,L_{p}},
\label{eq:u_average}
\end{equation}
where $L_{p}$ is the length of the pore (see
Fig.\,\ref{fig:schem_force_balance}) and $\Delta{P}$ is the
difference in static pressure from the bottom of the pore to the top
of the channel.    It is the critical value of this pressure that
determines the critical permeation condition for the oil droplet.

\vskip 0.05in

After substituting the average velocity into
Eq.\,(\ref{eq:drag_force}), one can obtain the following equation
for the drag force in terms of the pressure gradient along the pore:
\begin{equation}
F_{D}\propto\,\frac{f(\lambda)\,{W_{p}}^{2}\,|\Delta{P}|\,r_{d}}{L_{p}}.
\label{eq:drag_force_final}
\end{equation}
This force is counter-balanced by the capillary forces. Since the
droplet permeation due to the transmembrane pressure is initiated by
the displacement of the interface along the vertical direction, only
the component of forces along the vertical ($x$) direction needs to
be considered. Forces in other directions cancel each other due to
symmetry. Figure\,\ref{fig:schem_force_balance} shows the control
volume and the surface tension forces acting on the droplet. Note
that the surface tension forces act in the tangential directions at
the intersection of the dashed rectangular control volume and the
droplet surface. The $x$ component of the first force is estimated
$F_{1_{x}} \approx 4\,r_{d}\,\sigma\,\textrm{cos}\,\beta_{1}$ and
the second force acting on the droplet inside the channel is
$F_{2_{x}} \approx 4\,r_{d}\,\sigma\,\textrm{cos}\,\beta_{2}$.
The parameter $\sigma$ is the surface tension coefficient between
oil and water.

\vskip 0.05in

% Q: What is the physical significance of the angles beta1 and beta2?

By definition, the angle $\beta_{1}=180^{\circ}-\,\theta$, where
$\theta$ is the contact angle measured in oil, while the angle
$\beta_{2}$ can be approximated as
$\textrm{cos}\,\beta_{2}\approx\,\frac{W_p}{2\,r_d}$.  The model
assumes that the retentive capillary force is a result of the
competition between the capillary forces from two curved interfaces.
If the droplet size is much larger than the pore width, then
$\beta_{2}$ is close to $90^{\circ}$ and the dominant capillary
force will be at the oil-water interface inside the pore.

% Decreasing the size of the droplet reduces the difference between
% the two curvatures, and, therefore, the resultant capillary force is
% smaller and so is the drag force required to permeate the droplet.

\vskip 0.05in

\begin{figure}[t]
\includegraphics[width=13.0cm,angle=0]{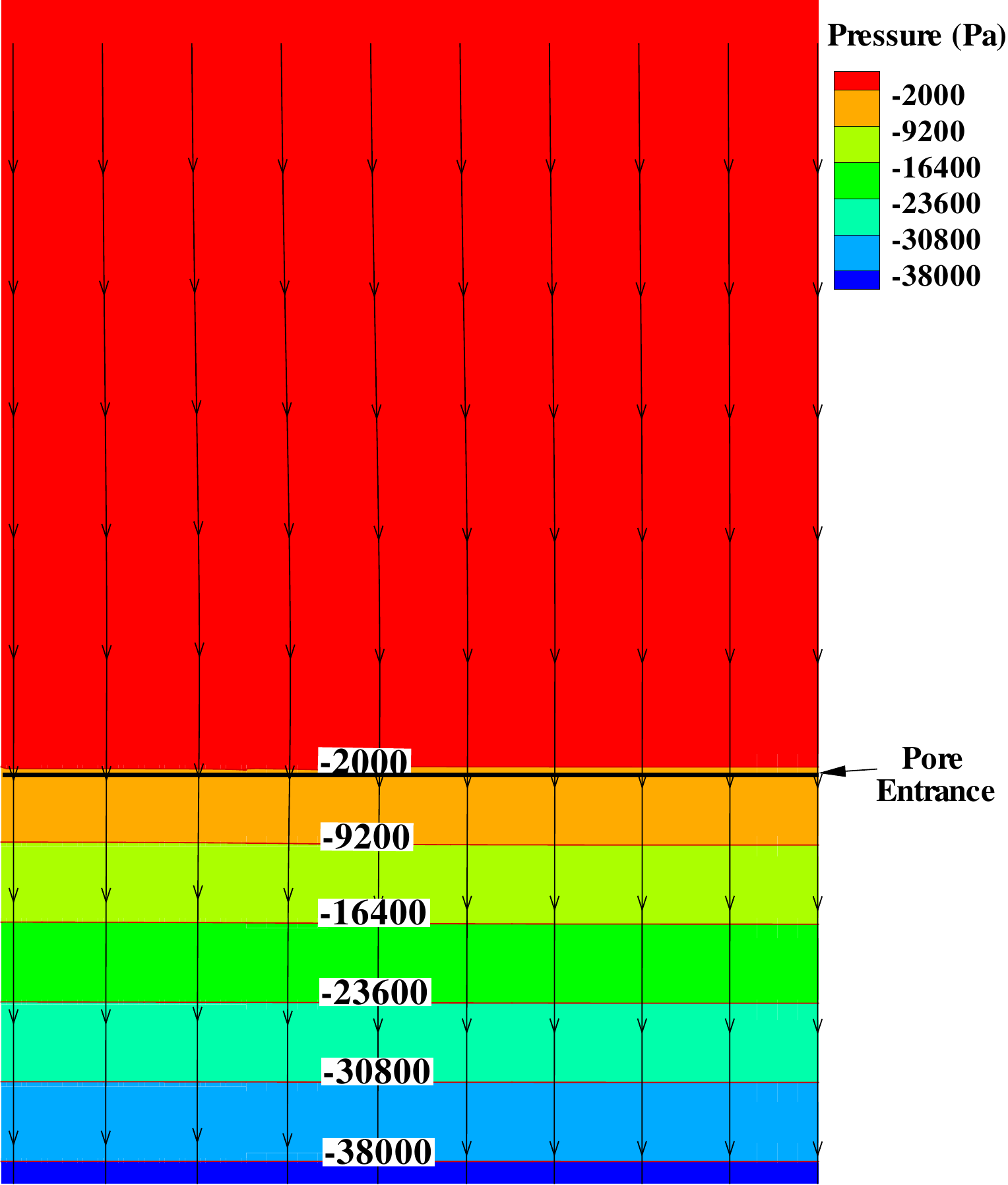}
\caption{(Color online) The pressure contours and streamlines along
the plane that crosses the simulation domain and the slotted pore.
The transmembrane pressure is $40\,\text{kPa}$. In the absence of an
oil droplet, the flow across the pore is uniform. Notice that
pressure along the streamlines varies linearly inside the pore. }
\label{fig:pressure_contours}
\end{figure}

The summation of the two forces along the $x$ direction yields:
\begin{equation}
F_{\sigma_{x}}\approx\,2\,\sigma\,W_{p}\,\Big(\frac{2\,r_{d}\,\textrm{cos}\,\beta_{1}}
{W_{p}}-1\Big).
\label{eq:surface_tension_force}
\end{equation}
At the critical permeation pressure, the forces in
Eq.\,(\ref{eq:drag_force_final}) and
Eq.\,(\ref{eq:surface_tension_force}) should be equal,
\textit{i.e.}, $F_{D}=F_{\sigma_{x}}$. Rearranging the terms gives:
%
% P_{cr}\propto\,\sigma\,(2\,\frac{r_{d}}{W_{p}}\,cos\beta_{1}-1)\,L_{p}/(f(\lambda)\,W_{p}\,r_{d})
% P_{cr}\propto\,\frac{\sigma\,(2\,r_{d}\,cos\beta_{1}/\,W_{p}-1)\,L_{p}}{f(\lambda)\,W_{p}\,r_{d}}
%
%
\begin{equation}
\Delta{P}\,=\,K\,\frac{\sigma\,(2\,r_{d}\,\textrm{cos}\,\beta_{1}/\,W_{p}-1)\,L_{p}}
{f(\lambda)\,W_{p}\,r_{d}},
\label{eq:critical_pres_slotted}
\end{equation}
where $K$ is the constant of proportionality. This expression,
however, does not account for the pressure variation inside the
pore.  In the case of viscous flow inside the pore with no-slip
boundary conditions (Hagen-Poiseuille flow), the pressure decreases
linearly along the pore length, as shown for example in
Fig.\,\ref{fig:pressure_contours}.

% No-slip condition on the surface inside the pore modifies the
% distribution of pressure inside the channel.

\vskip 0.05in

As will be shown in Sec.\,\ref{subsec:effect_of_parameters},
sufficiently large transmembrane pressure results in the deformation
of the oil droplet at the pore entrance.   For sub-critical
transmembrane pressures (below the critical value), the droplet
remains at the pore entrance but it will be deformed by permeating
incrementally inside the pore as the transmembrane pressure
increases.  Therefore, the surface of the droplet inside the pore is
exposed to a pressure already reduced due to the linear pressure
drop inside the pore, as shown in Fig.\,\ref{fig:pressure_contours}.
This pressure is proportional to the droplet size, since larger
droplets penetrate deeper inside the pore at sub-critical
conditions. Therefore, the extra term for pressure can be
approximated as $-\frac{r_{d}}{L_{p}}\,\Delta{P}$. Adding this term
to the right-hand side of Eq.\,(\ref{eq:critical_pres_slotted})
gives:
\begin{equation}
\Delta{P}=\,K\,\frac{\sigma\,(2\,r_{d}\,\textrm{cos}\,\beta_{1}/\,W_{p}-1)\,L_{p}}
{f(\lambda)\,W_{p}\,r_{d}}-\,\frac{r_{d}}{L_{p}}\,\Delta{P}.
\label{eq:critical_pres_complete}
\end{equation}
Rearranging Eq.\,(\ref{eq:critical_pres_complete}) and assuming
$\Delta{P}=P_{cr}$, gives the following relation for the critical
pressure of permeation of a droplet into an infinitely long slotted
pore:
\begin{equation}
P_{cr}=\,K\,\frac{\sigma\,(2\,r_{d}\,\textrm{cos}\,\beta_{1}/\,W_{p}-1)\,L_{p}}
{f(\lambda)\,W_{p}\,r_{d}\,(1+\,r_{d}/\,L_{p})}.
\label{eq:critical_pres_final}
\end{equation}

\vskip 0.05in

% Q: What are the basic parameters?

In the next section, we investigate the influence of oil-to-water
viscosity ratio, surface tension, contact angle, and droplet size on
the critical pressure of permeation using numerical simulations and
compare the numerical results with the predictions of
Eq.\,(\ref{eq:critical_pres_final}).  Throughout the paper, the
following values are used, unless explicitly noted otherwise:
$r_{d}=2\,{\mu}\text{m}$, $W_{p}=1\,{\mu}\text{m}$,
$L_{p}=5\,{\mu}\text{m}$, $\lambda=1$, $\theta=135^{\circ}$, and
$\sigma=19.1\,\text{mN/m}$.

\subsection{Effect of physiochemical parameters on the critical pressure of permeation}
\label{subsec:effect_of_parameters}

% ---- Summary

In this section, the numerical simulations of an oil droplet
permeating into an infinitely long slotted pore are performed (using
the computational methods described in
Sec.\,\ref{sec:numerical_method}) and the results are compared to
the theoretical model for the critical permeation pressure,
Eq.\,(\ref{eq:critical_pres_final}), presented in
Sec.\,\ref{subsec:theory}.

\vskip 0.05in

% Q: What is your geometry like? Is this really your rd?

In our computational setup, we have chosen a pore geometry that most
closely resembles a slotted pore with a very large cross-sectional
aspect ratio.  More specifically, the pore is modeled as a slit with
an infinite aspect ratio (which is implemented numerically by using
the symmetry boundary conditions).  This ensures that the oil
droplet is unaffected by the finite size effects and, possibly, by
other droplets on the same pore.  The pore is assumed to be a
perfectly rectangular cuboid with sharp edges. The value of the
droplet radius in our simulations is the radius of a spherical
droplet before depositing on the surface of the membrane. The radius
of the droplet after depositing at the pore entrance, which is used
in deriving the critical pressure given by
Eq.\,(\ref{eq:critical_pres_final}), is only slightly different from
the initial droplet radius and it is mainly dependent on the contact
angle.

\vskip 0.05in

% ---- Pressure distribution and streamlines

\begin{figure}[t]
\includegraphics[width=13.0cm,angle=0]{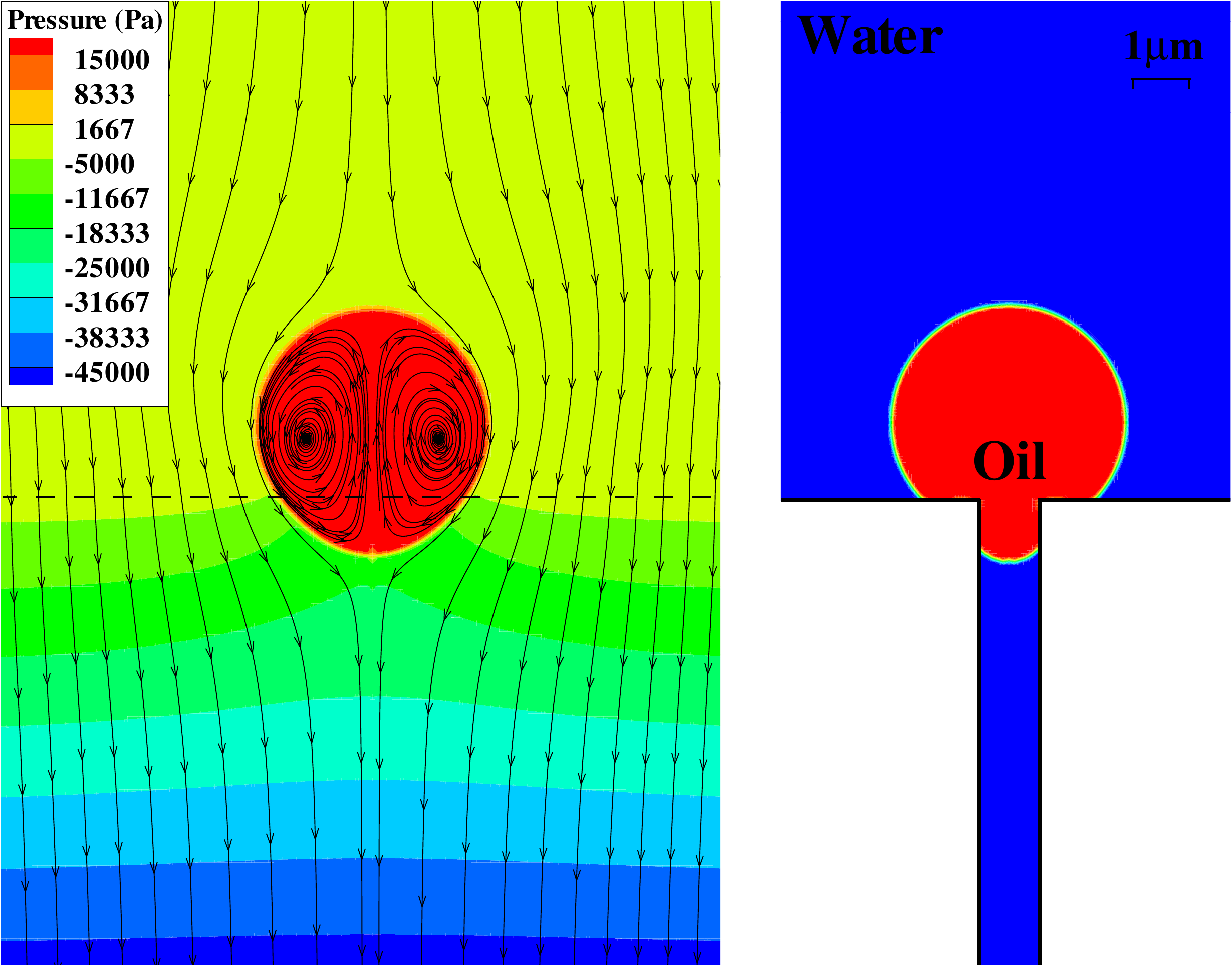}
\caption{(Color online) The left panel shows the distribution of
pressure and flow streamlines at the transmembrane pressure of
$45\,\text{kPa}$ (front view) and the right panel displays the oil
droplet at the entrance of the slotted pore (side view). The
horizontal dashed line on the left panel indicates the pore
entrance. }
\label{fig:contours}
\end{figure}

Figure\,\ref{fig:contours} shows the flow streamlines as well as
pressure distribution inside the pore and around the droplet at the
initial stage of permeation. It can be seen that the pressure inside
the droplet is higher than the outside pressure due to capillary
forces. In addition, the flow inside the droplet undergoes a
circulation caused by the shear stresses that are applied by the
flow at the interface.   As evident from Fig.\,\ref{fig:contours},
the streamlines begin at the top of the channel, continue across the
pore and exit at the bottom of the pore.  Notice that the flow is
curved around the droplet, and the pressure distribution below the
droplet deviates from the linear profile.  As the Reynolds number is
small, the flow is laminar and it remains attached to the droplet
surface.

\vskip 0.05in

% ---- Time evolution of permeation

\begin{figure}[t]
\includegraphics[width=8.0cm,angle=0]{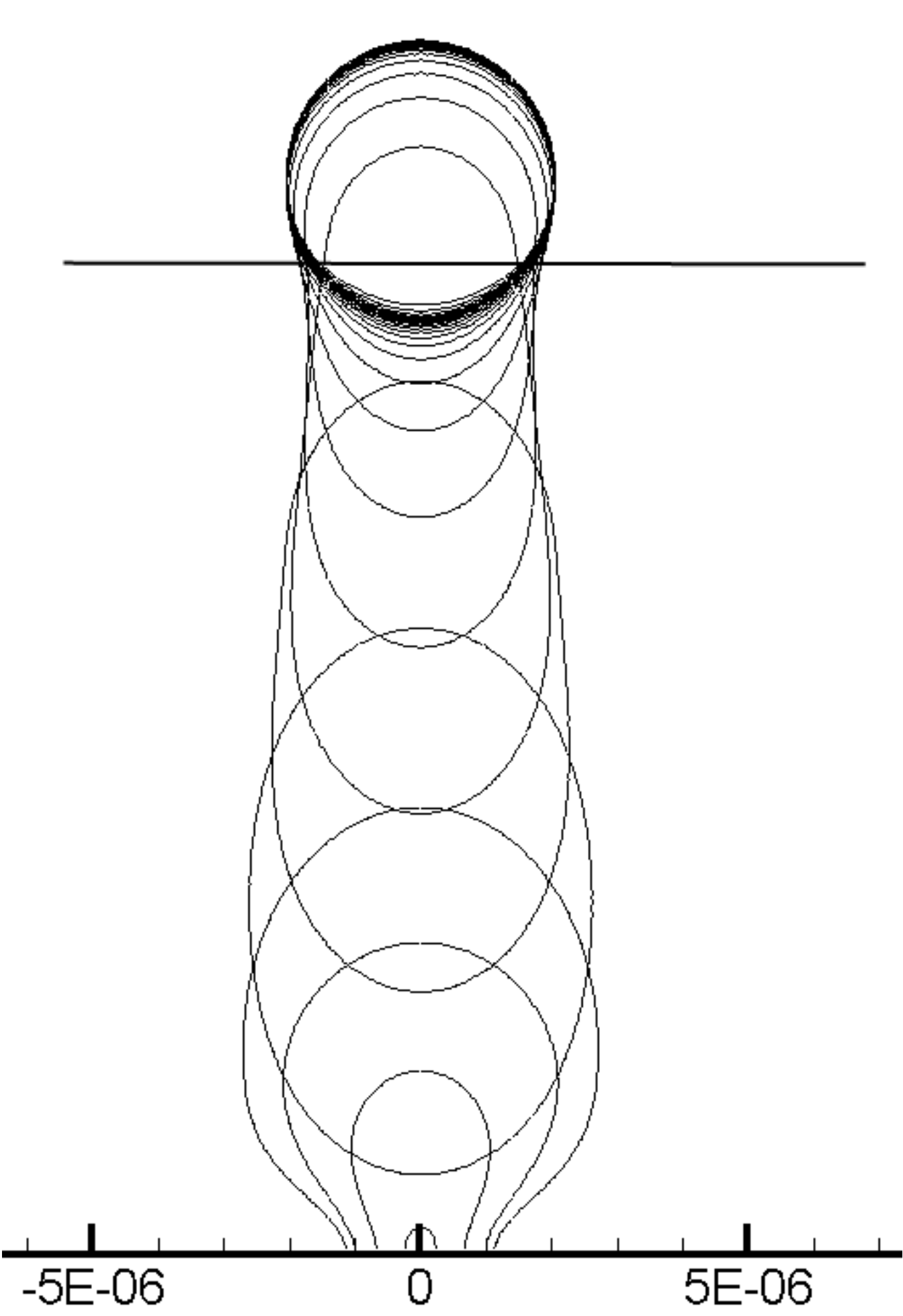}
\caption{The time sequence of the shape profiles inside the slotted
pore illustrating the droplet permeation process (see text for
details). The droplet radius is $r_{d}=2\,{\mu}\text{m}$. }
\label{fig:permeation_evolution}
\end{figure}

Figure\,\ref{fig:permeation_evolution} illustrates the shape
evolution of the oil droplet inside the slotted pore. In our setup,
the transmembrane pressure is initially set to a sub-critical value
and then gradually increased up to the critical pressure.  In the
sub-critical regime, the oil droplet remains at the pore entrance
and its lower part partially penetrates into the pore. When the
pressure is increased up to the critical value, the permeation
process accelerates drastically.  At first, the lower part of the
droplet starts stretching inside the pore, while the upper part
remains temporarily pinned.  Next, the droplet permeation is
accelerated as the bottom of the droplet inside the pore experiences
increasingly lower pressure as compared to the top of the droplet
inside the channel (see Fig.\,\ref{fig:pressure_contours}).
Eventually, the whole droplet enters the pore and deforms into an
oblate disc with the major axis oriented along the flow direction.
This stretching is mainly caused by the linear pressure gradient
inside the pore, as shown in Fig.\,\ref{fig:pressure_contours}. Once
the droplet is completely inside the pore, the permeation proceeds
very quickly. Finally, the droplet reaches the bottom of the pore
and exits from it.

\vskip 0.05in

% ---- Viscosity ratio

\begin{figure}[t]
\includegraphics[width=12.0cm,angle=0]{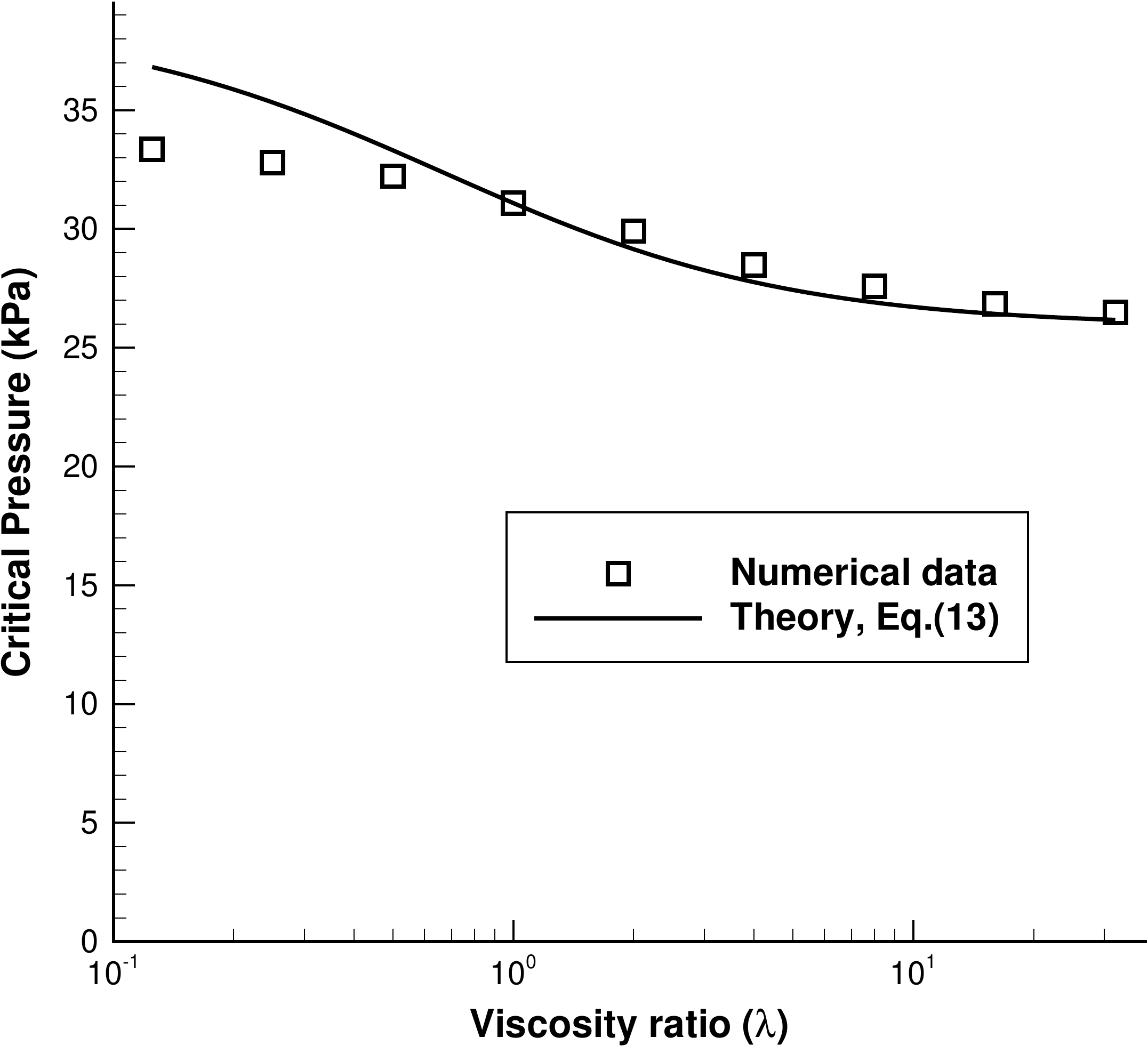}
\caption{The dependence of the critical permeation pressure as a
function of the oil-to-water viscosity ratio, $\lambda$.  Square
symbols denote the numerical results and the solid line is the
prediction of Eq.\,(\ref{eq:critical_pres_final}). The droplet
radius is $r_{d}=\,2\,{\mu}\text{m}$, the surface tension
coefficient is $\sigma=19.1\,\text{mN/m}$, the contact angle is
$\theta=135^{\circ}$, and the slotted pore dimensions are
$W_{p}=\,1\,{\mu}\text{m}$ and $L_{p}=\,5\,{\mu}\text{m}$. }
\label{fig:pcr_lambda}
\end{figure}

The effect of the oil-to-water viscosity ratio on the critical
pressure of permeation of an oil droplet into a slotted pore is
shown in Fig.\,\ref{fig:pcr_lambda}.   The results of numerical
simulations (indicated by square symbols in
Fig.\,\ref{fig:pcr_lambda}) demonstrate that the critical pressure
decreases monotonically with increasing viscosity ratio.   This is
because a larger viscosity ratio results in higher shear stress on
the droplet, which in turn leads to larger deformation and
consequently permeation at lower transmembrane pressures.   In our
analysis of the critical pressure,
Eq.\,(\ref{eq:critical_pres_final}), the function $f(\lambda)$ is
assumed to be same as for a spherical droplet in uniform viscous
flow and it is given by~\cite{Loth08}:
\begin{equation}
f(\lambda)=\,\frac{3\,\lambda+\,2}{\lambda+\,1}.
\label{eq:lambda_function}
\end{equation}
Furthermore, the proportionality constant in
Eq.\,(\ref{eq:critical_pres_final}) was chosen to fit the numerical
value of $P_{cr}$ at $\lambda=\,1$.  Thus, the solid curve in
Fig.\,\ref{fig:pcr_lambda} represents the theoretical prediction of
Eq.\,(\ref{eq:critical_pres_final}).   It can be seen that there is
a good agreement between the theoretical prediction and the
numerical results, especially for larger values of the viscosity
ratio.   With further increasing viscosity ratio, $f(\lambda)
\approx 3$ and the critical pressure approaches an asymptotic value,
as predicted by Eq.\,(\ref{eq:critical_pres_final}). The critical
pressure decreases by about $15\%$ from $\lambda=1$ to $\lambda=32$.
Therefore, it can be concluded that the critical permeation pressure
is a relatively weak function of the viscosity ratio as compared to
other parameters such as the surface tension coefficient (discussed
below).

% Based on Eq.\,(\ref{eq:critical_pres_final}), the critical pressure
% of permeation is inversely proportional to Eq.\,(\ref{eq:lambda_function}).
% Since all parameters other than viscosity ratio are constant, they
% could all be combined into one proportionality constant. We then use
% the numerically-obtained value of the critical pressure at
% $\lambda=\,1$ and computed this proportionality constant. Using the
% computed proportionality constant, the critical pressure is plotted
% in terms of the viscosity ratio as shown in Fig.\,\ref{fig:pcr_lambda}.

\vskip 0.05in

% ---- surface tension coefficient

\begin{figure}[t]
\includegraphics[width=12.0cm,angle=0]{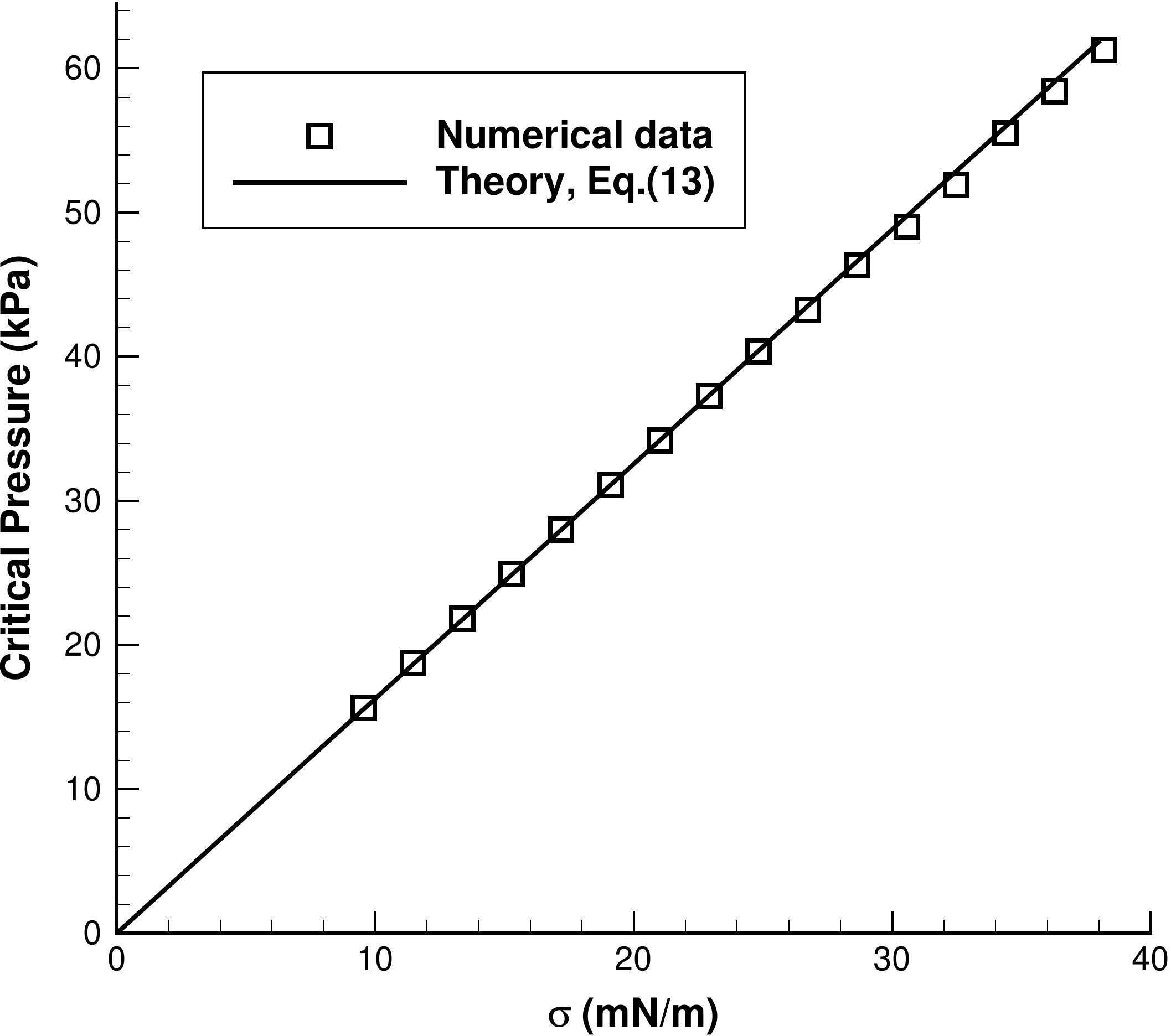}
\caption{The critical pressure of permeation of an oil droplet into
a slotted pore as a function of the surface tension coefficient. The
numerical results are indicated by symbols and the theoretical
prediction of Eq.\,(\ref{eq:critical_pres_final}) is denoted by the
solid line. }
\label{fig:pcr_sigma}
\end{figure}

We next consider the influence of surface tension on the critical
pressure of permeation of an oil droplet deposited on a slotted
pore.   The numerical results and theoretical prediction of
Eq.\,(\ref{eq:critical_pres_final}) are reported in
Fig.\,\ref{fig:pcr_sigma}.   Similar to the case of the viscosity
ratio, the proportionality coefficient in
Eq.\,(\ref{eq:critical_pres_final}) was gauged at the numerical
value of $P_{cr}$ at $\sigma=19.1\,\text{mN/m}$.  As is evident, the
critical pressure is a linear function of the surface tension
coefficient, implying that droplets with higher surface tension
require higher transmembrane pressures to enter the pore. The
excellent agreement between numerical results and analytical
predictions also confirms the linearity of surface tension effect in
the force balance analysis presented in Sec.\,\ref{subsec:theory}.
The fact that the extended theoretical line passes through zero
critical pressure demonstrates that surface tension is one of the
main factors that determine the permeation of the oil phase through
porous filters.

\vskip 0.05in

% ---- Contact angle

\begin{figure}[t]
\includegraphics[width=12.0cm,angle=0]{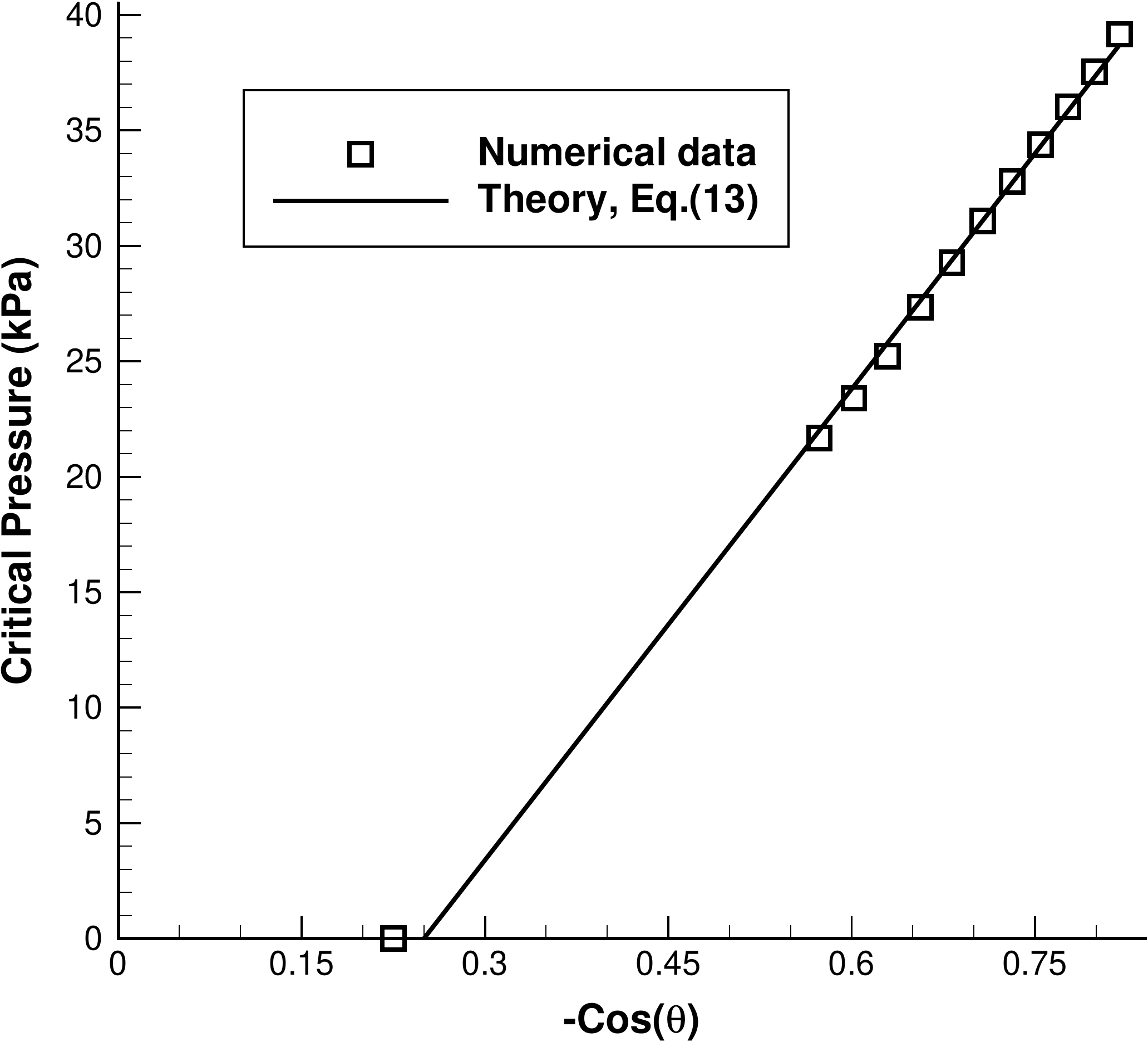}
\caption{The variation of the critical permeation pressure versus
$-\textrm{cos}\,\theta$. The numerical results and analytical
prediction of Eq.\,(\ref{eq:critical_pres_final}) are indicated by
symbols and the solid line, respectively.   The square symbol at
$-\textrm{cos}\,\theta=0.23$ denotes the numerical value of the
critical contact angle when $P_{cr}=0$. }
\label{fig:pcr_theta}
\end{figure}

Figure\,\ref{fig:pcr_theta} shows the critical pressure of
permeation as a function of the factor $-\textrm{cos}\,\theta$. To
remind, $\theta$ is the contact angle measured in oil and it is
related to the angle $\beta_{1}=180^{\circ}-\,\theta$ used in
Eq.\,(\ref{eq:critical_pres_final}) and shown in
Fig.\,\ref{fig:schem_force_balance}.  The theoretical line is fitted
to the numerical data using Eq.\,(\ref{eq:critical_pres_final}) at
the contact angle $\theta=135^{\circ}$, which corresponds to
$-\textrm{cos}\,\theta=0.707$, similar to the cases of viscosity
ratio and surface tension.   As seen in Fig.\,\ref{fig:pcr_theta},
the theoretical predictions agree well with numerical results,
indicating a nearly linear dependence of the critical pressure on
$-\textrm{cos}\,\theta$.   As discussed above, permeation of the
droplet into the pore results in an increase in the effective
surface area between oil and the pore wall (see
Fig.\,\ref{fig:permeation_evolution}). Thus, with increasing contact
angle, the surface energy between the oil phase and the pore surface
increases, and, as a result, the critical transmembrane pressure
becomes higher.

\vskip 0.05in

% ---- Inference of the critical contact angle for zero pressure:

Furthermore, it can be inferred from
Eq.\,(\ref{eq:critical_pres_final}) that the critical permeation
pressure becomes negative if $\textrm{cos}\,\beta_{1}<
W_{p}/2\,r_{d}$. This also implies that at the contact angle of
$90^{\circ}$, the droplet will permeate into the pore even at zero
transmembrane pressure.  Therefore, there should be a minimum
contact angle, above which the droplet will remain at the membrane
surface when there is no transmembrane pressure applied. This
critical contact angle can be readily estimated from
Eq.\,(\ref{eq:critical_pres_final}) by equating the right-hand side
to zero and using the definition $\beta_{1}=180^{\circ}-\,\theta$ to
obtain:
\begin{equation}
\theta_{cr}=\,180^{\circ}-\,\arccos\Big(\frac{W_{p}}{2\,r_{d}}\Big).
\label{eq:critical_theta}
\end{equation}
Plugging in our standard parameters $r_{d}=2\,{\mu}\text{m}$ and
$W_{p}=1\,{\mu}\text{m}$ into Eq.\,(\ref{eq:critical_theta}) results
in the critical contact angle of $\theta=104.5^{\circ}$. This value
is confirmed by numerical simulations to be $\theta=103^{\circ}$, as
shown in Fig.\,\ref{fig:pcr_theta} by the square symbol at
$-\textrm{cos}\,\theta=0.23$.

\vskip 0.05in

% Droplet size

\begin{figure}[t]
\includegraphics[width=12.0cm,angle=0]{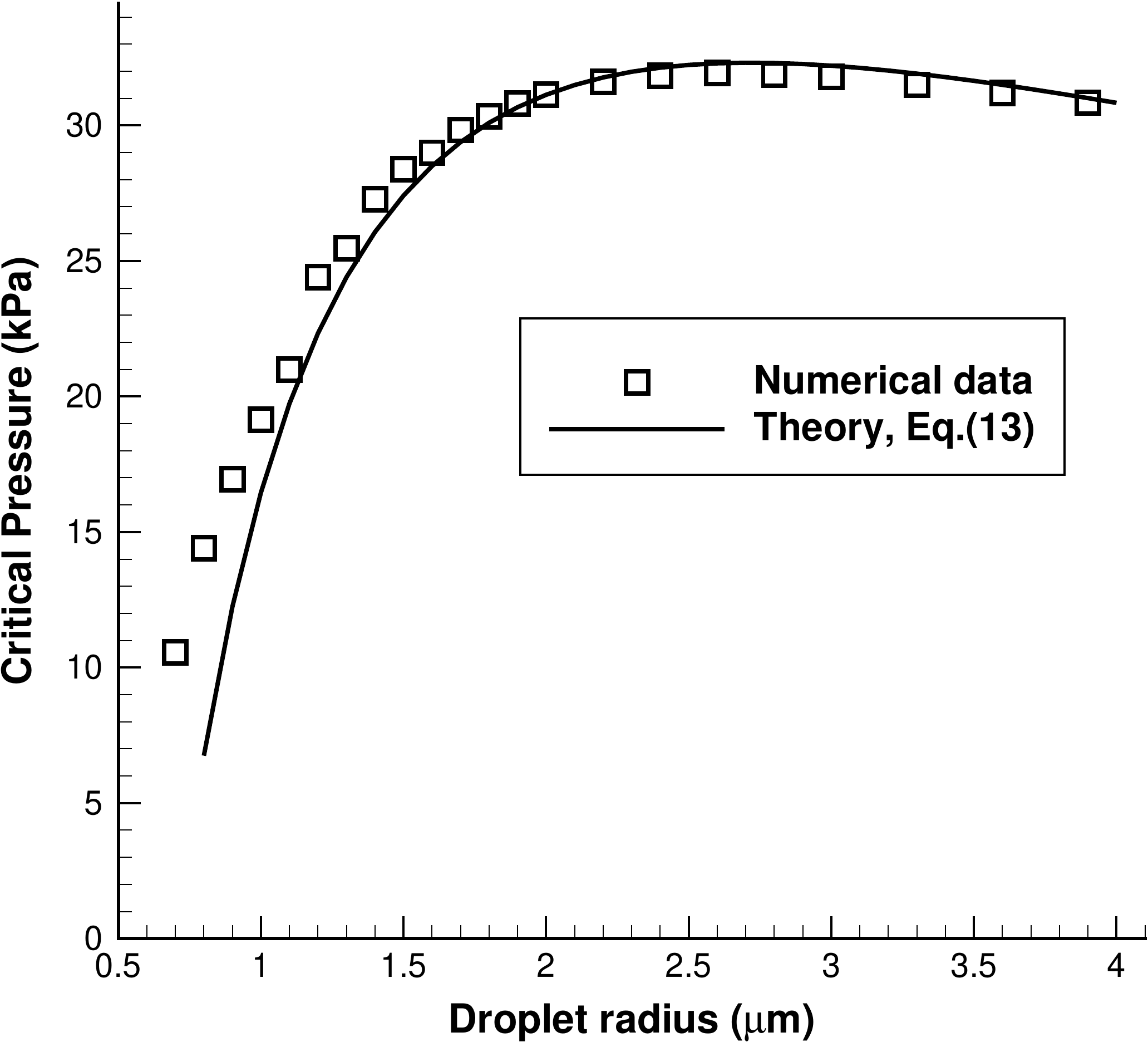}
\caption{The critical permeation pressure as a function of the
droplet radius. The simulation results are denoted by square
symbols, while the theoretical prediction of
Eq.\,(\ref{eq:critical_pres_final}) is indicated by the solid
curve.}
\label{fig:pcr_rd}
\end{figure}

The size of the oil droplet has two effects on the balance of forces
that determine the critical pressure of permeation.  First, as
predicted by Eq.\,(\ref{eq:drag_force_final}), larger droplets
experience higher drag force from the flow passing around its
surface. Second, as discussed in Sec.\,\ref{subsec:theory}, larger
droplets tend to penetrate deeper into the pore even at sub-critical
pressures. This results in an extra pressure gradient in the flow
direction and, thus, indirectly contributes to the drag force as
well [\,see Eq.\,(\ref{eq:critical_pres_complete})\,]. The results
of numerical simulations as well as the theoretical prediction of
Eq.\,(\ref{eq:critical_pres_final}) are summarized in
Fig.\,\ref{fig:pcr_rd}.   The theoretical curve is fitted to
$P_{cr}$ at $r_{d}=2\,{\mu}\text{m}$ using
Eq.\,(\ref{eq:critical_pres_final}), similar to the previous cases.
One can observe a good agreement between the theoretical curve and
the numerical data.  The results in Fig.\,\ref{fig:pcr_rd} show that
the critical pressure of permeation increases with the droplet
radius up to a certain value.   In particular, it can be seen that a
droplet with the radius of about $2.4\,{\mu}\text{m}$ requires the
highest permeation pressure. This trend can be understood from the
fact that the drag force due to the extra pressure gradient is
smaller than the surface tension forces, and, therefore, as the
droplet size increases, so does the transmembrane pressure needed
for the droplet permeation. However, for droplets larger than a
certain size, the drag force caused by the extra pressure gradient
becomes relatively large, and, as a result, the critical permeation
pressure becomes only weakly dependent on the droplet size (see
Fig.\,\ref{fig:pcr_rd}).

% *** Inference for the maximum drop size:

% Based on Eq. (), as the drop size increases, the critical pressure
% increases until it reaches a maximum value and it decreases
% afterwards. If the drop size is much higher than the pore size, the
% second term in Eq. () has a stronger effect than the first term and
% thus result in the reduction of the critical pressure.
% Equation () has a maximum value which could be evaluated by equating
% the numerator of its derivative to zero:
%
% \begin{equation}
% \frac{2\,cos\beta_{1}}{W_{p}\,L_{p}}\,r^{2}_{d}+\frac{2}{L_{p}}\,r_{d}+\,1=\,0
% \label{eq:deriv}
% \end{equation}
%
% The negative root of Eq. () is not a physical solution. The positive
% root could be simplified as:
%
% \begin{equation}
% r_{d_{max}}=\,W_{p}\times\,(\frac{1+\,\sqrt{1+\,2\,L_{p}\,cos\beta_{1}/\,W_{p}}}{2\,cos\beta_{1}})
% \label{eq:rd_max}
% \end{equation}
%
% This is a strong function of $L_{p}$ indicating that the longer the
% pore, the higher the value of drop radius that will maximize Eq. ().
% This is because the effect of the pressure gradient inside the pore
% is less pronounced for longer pores.

\section{Conclusions}

In summary, we investigated numerically the problem of an oil
droplet permeation into a slotted pore under the applied
transmembrane pressure. This situation is relevant to filtration of
dilute oil-in-water emulsions, where the interaction between
different droplets is not important and a single oil droplet blocks
only a part of the slotted pore, thus allowing a finite permeate
flux.  More specifically, we considered an infinitely long slotted
pore with sharp edges and parallel, nonwetting interior walls. The
critical permeation pressure was first estimated theoretically by
considering a difference between a drag force due to flow around a
droplet and capillary forces due to curved oil-water interfaces, and
taking into account a finite pressure variation inside the pore.
The results of numerical simulations demonstrated that the critical
pressure as a function of the oil-water viscosity ratio, surface
tension, contact angle and droplet size agree well with theoretical
predictions.  In addition, an estimate of the critical contact
angle, required for the permeation at zero applied pressure, was
obtained and verified numerically.   The simulation results also
suggest that above a certain droplet size, the critical permeation
pressure is only weakly dependent on the droplet radius.

\vskip 0.05in

Altogether, the numerical analysis of the droplet dynamics at the
pore entrance and the theoretical prediction for the critical
permeation pressure might be useful for development of efficient
microfiltration systems. For example, by choosing an appropriate
pressure across a slotted pore membrane, a dilute monodisperse
oil-water emulsion that consists of two different dispersed fluids
can be separated based on the viscosity ratio difference, thus
allowing only one of the dispersed fluids to permeate. It is
important to reiterate, however, that the key parameters that
strongly influence the critical permeation pressure include the
contact angle, surface tension, and drop-to-pore ratio.

\section*{Acknowledgments}

Financial support from the National Science Foundation (CNS-1531923
and CBET-1033662) and the Michigan State University Foundation
(Strategic Partnership Grant 71-1624) is gratefully acknowledged.
The article was prepared within the framework of the Basic Research
Program at the National Research University Higher School of
Economics (HSE) and supported within the framework of a subsidy by
the Russian Academic Excellence Project `5-100'. Computational work
in support of this research was performed at Michigan State
University and Wright State University Computing Facilities and the
Ohio Supercomputer Center.

%%%%%%%%%%%%%%% FIGURES %%%%%%%%%%%%%%%%%%%%%%%

% FIG 1

% FIG 2

% FIG 3

% FIG 4

% FIG 5

% FIG 6

% FIG 7

% FIG 8

\bibliographystyle{prsty}

\end{document}